\documentclass[showpacs,aps,prb,amsmath,amssymb]{revtex4}
\usepackage{graphicx}
\usepackage{dcolumn}
\usepackage{bm}
\begin{document}

\renewcommand{\thefigure}{\arabic{figure}}
\title{Correlation energy of a two-dimensional electron gas from static and dynamic 
exchange-correlation kernels}
\author{R. Asgari,$^{1,2}$ M. Polini,$^{1,3}$ B. Davoudi,$^{1,2}$ and M. P. Tosi$^{1}$}
\affiliation{$^1$NEST-INFM and Classe di Scienze, Scuola Normale Superiore, I-56126 Pisa, Italy\\
$^2$Institute for Studies in Theoretical Physics and Mathematics, Tehran
19395-5531, Iran\\
$^3$Department of Physics, The University of Texas at Austin, Austin, Texas 78712, USA
}
\begin{abstract}
We calculate the correlation energy of a two-dimensional homogeneous electron gas 
using several available approximations for the exchange-correlation kernel $f_{\rm xc}(q,\omega)$ 
entering the linear dielectric response of the system. As in the previous work of Lein {\it et al.} [Phys. Rev. 
B {\bf 67}, 13431 (2000)] on the three-dimensional electron gas, we give attention to the relative 
roles of the wave number and frequency dependence of the kernel and analyze the correlation 
energy in terms of contributions from the $(q, i\omega)$ plane. We find that consistency of the kernel 
with the electron-pair distribution function is important and in this case the nonlocality of the 
kernel in time is of minor importance, as far as the correlation energy is concerned. We also 
show that, and explain why, the popular Adiabatic Local Density Approximation performs 
much better in the two-dimensional case than in the three-dimensional one.
\end{abstract}
\maketitle
\section{Introduction}
Fluids of electronic carriers with essentially two-dimensional (2D) dynamics in semiconductor structures present a rich phenomenology, in which dynamical correlations take increasing importance as the electron density is lowered\cite{1}. Many of the electron-electron interaction effects can be understood with the help of the homogeneous electron gas (EG) model\cite{2}. 
A central role in the theory of short-range correlations in the EG is played by the wave 
number and frequency dependent exchange-correlation kernels or equivalently by the local field 
factors entering the linear response properties of the model\cite{3}. These provide key inputs for 
some applications of density functional theory\cite{4} and for studies of quasi-particle properties 
such as the effective mass and the effective Land\`e $g$-factor\cite{5}. These properties are known 
from experiment for carriers in semiconductor structures over a wide range of carrier density\cite{6}. 
A great deal of accurate information has come from quantum Monte Carlo (QMC) 
studies\cite{7}, but theoretical understanding continues to draw interest.

The exchange-correlation kernel $f_{\rm xc}(q,\omega)$ entering the dielectric response of the EG 
determines its correlation energy through the fluctuation-dissipation theorem relating the 
imaginary part of the inverse dielectric function to van Hove's dynamic structure factor $S(q,\omega)$. 
Starting from $S(q,\omega)$ an integration over frequency is needed to obtain the electron-pair 
structure factor $S(q)$, which in Fourier transform describes the shape of the instantaneous 
exchange-correlation (Pauli-Coulomb) hole surrounding an average electron located at the 
origin. From $S(q)$ the ground-state energy $E_g$ can be calculated by means of further integrations 
over wave number and over coupling strength (see for instance Ref. \onlinecite{3}) and the correlation energy $\varepsilon_c$ is defined, as usual, as the difference between $E_g$ and its Hartree-Fock value.
	
For the 3D EG Lein {\it et al.}\cite{8} have analyzed in detail this procedure for calculating the 
correlation energy, with particular attention to the roles of the wave number and frequency 
dependence of $f_{\rm xc}(q,\omega)$. While the wave number dependence of the kernel is physically related 
to the spatial shape of the exchange-correlation hole, as already remarked, its frequency 
dependence reflects the inertia of the hole as the electron at the origin moves through the EG. 
When the frequency dependence of $f_{\rm xc}(q,\omega)$ is omitted and its wave number dependence is 
approximated by its leading long-wavelength term, one is accounting for short-range exchange 
and correlation effects in the EG only through the compressibility sum rule (see for instance Ref. \onlinecite{3}). This approximation provides the basis for the so-called Adiabatic Local Density Approximation (ALDA), which has often been used in dealing with time-dependent phenomena in inhomogeneous 
electronic systems in the low-frequency regime\cite{4}.

In the present work we extend the analysis of Lein {\it et al.}\cite{8} to the 2D EG by examining 
how several known forms of $f_{\rm xc}(q,\omega)$ perform in regard to the calculation of its correlation 
energy over a range of values for the coupling strength. The standard of comparison is the 
parametrized form reported for the correlation energy by Rapisarda and Senatore\cite{9} from their 
Diffusion QMC results. The paper is organized as follows. In Sec. II 
we report the detailed expressions relating $\varepsilon_c$  to $f_{\rm xc}(q,\omega)$ 
and in Sec. III we list the various forms of $f_{\rm xc}(q,\omega)$ that we have considered. 
Section IV presents our main numerical results and 
discusses them with special emphasis on the usefulness of the ALDA in 2D as compared with 
3D. Finally, in an Appendix we briefly comment on the scaling properties of the exchange-correlation 
kernel for inhomogeneous electron fluids of arbitrary dimensionality.
\section{Correlation energy from exchange-correlation kernels}
We consider a fluid of electrons moving in a plane 
and interacting by the $e^2/r$ law. The 
correlation energy (per electron) can be written in terms 
of the difference between the potential 
energy of the fluid and the exchange energy, integrated over the 
coupling-strength parameter $\lambda$:
\begin{equation}\label{1}
\varepsilon_c=\frac{1}{2}\int_0^{e^2} \frac{d \lambda}{\lambda} \int \frac{d^{2} {\bf q}}{(2\pi)^{2}}\, 
v^{\lambda}_{q}[S_{\lambda}(q)-S_{\rm \scriptstyle HF}(q)]\,.
\end{equation}
Here $v^{\lambda}_{q}=2 \pi \lambda/q$ is the 2D Fourier 
transform of the Coulomb potential with strength $\lambda$, 
$S_{\lambda}(q)$ is the structure factor of an EG with interactions 
$v^{\lambda}_{q}$ and $S_{\rm \scriptstyle HF}(q)$ 
is the Hartree-Fock structure factor,
\begin{equation}\label{2}
S_{\rm \scriptstyle HF}(q)=\left\{
\begin{array}{l}
\frac{2}{\pi}\left[\arcsin{({\bar q})}+{\bar q}\,\sqrt{1-{\bar q}^2}\right]\hspace{0.6 cm}\mbox{for ${\bar q}< 1$}\\
1\hspace{4.47 cm}\mbox{for ${\bar q}> 1$}
\end{array}
\right.
\end{equation}
where ${\bar q}=q/(2 k_F)$ with $k_F=(2 \pi n)^{1/2}$ being the Fermi wave number determined 
by the 2D electron density $n$. The fluctuation-dissipation theorem relates the structure factor 
to the density-density response function 
$\chi^{\lambda}_{\rho \rho}(q,i u)$ calculated on the imaginary frequency axis,
\begin{equation}\label{3}
S_{\lambda}(q)=-\frac{\hbar}{\pi n}\int_{0}^{\infty}d u\,\chi^{\lambda}_{\rho \rho}(q,i u)\,.
\end{equation}
The response function 
can in turn be expressed through the frequency-dependent exchange-correlation kernel 
$f_{\rm xc}(q,\omega)$ on the imaginary 
frequency axis,
\begin{equation}\label{4}
\chi^{\lambda}_{\rho \rho}(q,i u)
=\frac{\chi_0(q,i u)}{1-[v^{\lambda}_{q}+f^{\lambda}_{\rm xc}(q, i u)] \chi_0(q,i u)}
\end{equation}
where
\begin{eqnarray}\label{5}
\chi_{0}(q,iu)=-\frac{m}{\pi \hbar^2} \left[\displaystyle 1-\frac{1}{\sqrt{2}\,{\bar q}}\,\sqrt{\displaystyle f({\bar q},{\bar u})+\sqrt{\displaystyle f^2({\bar q},{\bar u})+4{\bar q}^2{\bar u}^2}}~\right]
\end{eqnarray}
is the density-density response function of the 2D ideal Fermi gas\cite{10}. 
Here, ${\bar u}=m u/(\hbar k_{F}\,q)$ and 
$f({\bar q},{\bar u})={\bar q}^2-{\bar u}^2-1$.	

From Eqs. (3) and (4) it is possible to rewrite Eq. (1) in the form
\begin{equation}\label{6}
\varepsilon_c=-\frac{\hbar}{4 \pi^2 n}\int_0^{\infty} q d q\int_0^{\infty}d u \int_0^{e^2} \frac{d \lambda}{\lambda}\, v^{\lambda}_q\,\frac{[\chi_{0}(q,iu)]^2\left[v^{\lambda}_{q}+f^{\lambda}_{\rm xc}(q, i u)\right]}{1-[v^{\lambda}_{q}+f^{\lambda}_{\rm xc}(q, i u)] \chi_0(q,i u)}\, .
\end{equation}
In the paramagnetic case of present interest $\varepsilon_c$ depends only on $n$, or equivalently on the 
dimensionless density parameter $r_s$ defined by $r_s a_B=(\pi n)^{-1/2}$, $a_B$ being the Bohr radius.

In fact, the full dependence of $f^{\lambda}_{\rm xc}(q, i u)$ on $\lambda$ 
is not needed for calculating $\varepsilon_c$ at any given $r_s$. 
It suffices to know the $r_s$ dependence of the exchange-correlation kernel at full coupling 
strength, {\it i.e.} $f_{\rm xc}(q, i u)\equiv f^{\lambda=e^2}_{\rm xc}(q, i u)$. 
That is, the correlation energy can be calculated from
\begin{equation}\label{7}
\varepsilon_c(r_s)=\left[\frac{\sqrt{2}}{r_s^2}\int_0^{r_s}\gamma_c(r'_s)\,d r'_s\right]\, {\rm Ryd}
\end{equation}
where
\begin{equation}\label{8}
\gamma_c(r_s)=\frac{1}{k_F}\int_0^{\infty}d q\, [S(q)-S_{\rm \scriptstyle HF}(q)]=-\frac{\hbar }{\pi n k_F }
\int_0^{\infty}d q\int_0^{\infty}d u \,\frac{[\chi_{0}(q,iu)]^2\left[v_{q}+f_{\rm xc}(q, i u)\right]}{1-[v_{q}+f_{\rm xc}(q, i u)] \chi_0(q,i u)}\,.
\end{equation}
Several known approximations to $f_{\rm xc}(q, i u)$ as needed for this calculation will be recalled in the 
next Section.

\section{Approximate kernels for the homogeneous electron gas}
We have tested five main forms of the kernel $f_{\rm xc}(q, i u)$. 
In three of these (designated by the acronyms ALDA, DPGT, and STLS) 
the frequency dependence of the kernel is omitted. 
A dynamical kernel has been used in the approximations designated by QSTLS and AKA. Some 
details are as follows. 

(i) ALDA: for the homogeneous 
electron gas the adiabatic local density approximation omits 
both the wave number and the frequency dependence 
of the kernel\cite{4}. It amounts to taking into 
account the compressibility sum rule by setting
\begin{equation}\label{9}
f_{\rm xc}^{\rm \scriptstyle ALDA}(q, iu)=
\frac{1}{n^2\kappa_0}\left(1-\frac{\kappa_0}{\kappa}\right)
\end{equation}
where $\kappa$ is the compressibility of the EG and $\kappa_0$ that of the ideal Fermi gas, given in 2D by
$\kappa_{0}=\pi r^{4}_{s}/2$ in units of $a^{2}_{B}/{\rm Ryd}$. The ratio $\kappa_{0}/\kappa$ is related to the correlation energy (in Ryd units) by
\begin{equation}\label{10}
\frac{\kappa_{0}}{\kappa}=1-\frac{\sqrt{2}}{\pi}\,r_{s}+\frac{r^{4}_{s}}{8}
\left[\frac{d^{2} \epsilon_{c}}{d r^{2}_{s}}-\frac{1}{r_{s}}
\frac{d \epsilon_{c}}{d r_{s}}\right]\,.
\end{equation}
We evaluate this input from the parametrized form of $\varepsilon_{c}(r_s)$ of Rapisarda and Senatore\cite{9}. 

(ii) DPGT: an analytical expression for the {\it static} kernel 
$f_{\rm xc}(q, 0)$ as a function of $r_s$ in the 
range $0\leq r_s \leq 10$ has been obtained by Davoudi {\it et al.}\cite{11} by fitting the available Diffusion 
Monte Carlo data for the local field factor in the 2D EG\cite{12}. Their expression embodies the 
compressibility sum rule as well as the asymptotic high-$q$ behavior of the static kernel. By 
setting
\begin{equation}\label{11}
f_{\rm xc}^{\rm \scriptstyle DPGT}(q, iu)=f_{\rm xc}^{\rm \scriptstyle QMC}(q, 0)
\end{equation}
we are evidently omitting the frequency dependence of the kernel but taking into account its 
"exact" dependence on wave number at low frequency. 

(iii) STLS: the Singwi-Tosi-Land-Sj\"olander approximation\cite{13} omits the frequency 
dependence of the kernel by relating it self-consistently to the structure 
factor $S(q)$ through
\begin{equation}\label{12}
f_{\rm xc}^{\rm \scriptstyle STLS}(q, iu)=\frac{1}{n}\,\int\frac{d^2 {\bf k}}{(2\pi)^2}\,\frac{{\bf q}\cdot{\bf k}}{q^2}\,v_{k}\left[S(|{\bf q}-{\bf k}|)-1\right]\, .
\end{equation} 
This expression was justified by an 
analysis of kinetic equations in the presence of a time-
dependent weak external potential and has been used 
rather widely in the literature. 

(iv) QSTLS: a frequency dependence of the local field factor in an 
STLS-type self-consistent theory has been included by later authors. 
We adopt in particular the expression\cite{14}
\begin{equation}\label{13}
f_{\rm xc}^{\rm \scriptstyle QSTLS}(q, iu)=\frac{1}{n}\int\frac{d^2 {\bf k}}{(2\pi)^2}\, 
\frac{\chi_0({\bf q}, {\bf k}; i u)}{\chi_0(q,i u)}~v_{k}\left[S(|{\bf q}-{\bf k}|)-1\right]
\end{equation}
where the function $\chi_0({\bf q}, {\bf k}; i u)$ is given by\cite{15}
\begin{equation}\label{14}
\chi_0({\bf q}, {\bf k};i u)=-\frac{m}{\pi \hbar^2}\,c\,(1-r)\,.
\end{equation}
Here $c={\bf q}\cdot{\bf k}/q^2$ and
\begin{equation}\label{15}
r^2=-\frac{y^2+z^2-c^2}{2c^2}+\frac{\left[(y^2+z^2-c^2)^2+4c^2z^2\right]^{1/2}}{2c^2}
\end{equation}
with $y=2 k_F/ q$ and $z=2mu/\hbar q^2$. 

(v) AKA: Atwal {\it et al.}\cite{16} have recently proposed an analytic expression for the dynamical 
local field factor which is based on many-body perturbation theory and on known sum rules, in 
the same spirit as in the work of Richardson and Ashcroft\cite{17} on the 3D EG. With the 
notations of Ref. \onlinecite{16},
\begin{equation}\label{16}
f_{\rm xc}^{\rm \scriptstyle AKA}(q, iu)=-v_{q}\left[G_s(q, i u)+G_n(q, i u)\right]
\end{equation}
where
\begin{equation}\label{17}
G_s(q, i u)=\frac{a_s(u)q+b_s(u)q^7}{1+c_s(u)q+d_s(u)q^7}
\end{equation}
and
\begin{equation}\label{18}
G_n(q, i u)=\frac{a_n(u)q+b_n(u)q^7}{1+c_n(u)q+d_n(u)q^6}\,.
\end{equation}
Here $a_{s,n}(u)$, $b_{s,n}(u)$, $c_{s,n}(u)$ and $d_{s,n}(u)$ are functions of frequency and depend on two input 
parameters, which are the contact value $g(0)$ of the pair distribution function and the correlation 
energy. In our calculations we have replaced Eqs. (41) and (52) of Ref. \onlinecite{16} for $d_{s,n}(u)$, which 
contain typographical errors, by the following expressions communicated to us by Dr. G. S. 
Atwal:
\begin{equation}\label{19}
d_s(\omega)=\frac{\zeta_s \lambda^0_s}{6[\zeta_s \lambda^0_s-1+g(0)](1+\omega^4)}+\frac{\lambda^{\infty}_s\omega}{[1-g(0)](1+\omega^4)}\,,
\end{equation}
\begin{equation}\label{20}
d_n(\omega)=\frac{\zeta_n \lambda^0_n}{(5\zeta_n \lambda^0_n+2\lambda^{\infty}_n)(1+\omega^4)}+\frac{\gamma_n \omega^{9/2}}{[5\gamma_n\omega^{1/2}+2\lambda^{\infty}_n(1+0.38\omega)^{1/2}](1+0.38\omega)^3(1+\omega^4)}\,.
\end{equation}
For expressions of the other functions and of the various parameters the reader is referred to the 
original paper. 

In determining the input for this form of the kernel we have again used the form of 
Rapisarda and Senatore\cite{9} for $\varepsilon_c(r_s)$ 
and followed Atwal {\it et al.}\cite{16} 
in taking the values of $g(0)$ 
from an interpolation formula proposed by Polini {\it et al.}\cite{18}. We have also checked simple 
variants of this scheme, taking $g(0)$ from calculations in the ladder approximation\cite{19} or 
dropping the frequency dependence of $f_{\rm xc}^{\rm \scriptstyle AKA}(q, iu)$.

\section{Numerical results and discussion}
We turn to a presentation of our numerical results, which are collected in Figures 1-4. In 
Figure 1 (top panel) we show the difference between the correlation energy calculated by means 
of Eqs. (\ref{7}) and (\ref{8}) and the "exact" correlation energy $\varepsilon^{\rm \scriptscriptstyle QMC}_c$ 
from the parametrization of Ref. \onlinecite{9}, 
for electron densities in the range $0.5 \leq r_s \leq 10$. 
In 2D this range extends into the intermediate-to-strong coupling regime. 
In the bottom panel of Figure 1 we report the calculated correlation 
energies in the weak-coupling regime $0 \leq r_s \leq 0.8$, 
to show that none of the theories that we 
have examined satisfies the exactly known limiting expression\cite{20} for $r_s \rightarrow 0$,
\begin{equation}
\varepsilon_c(r_s)\rightarrow -0.385-\frac{2\sqrt{2}}{3\pi}(10-3\pi)r_s\ln{r_s}+o(r_s)\,.
\end{equation}
This panel also reports the correlation energy given by the Random Phase Approximation 
(RPA), which corresponds to setting the exchange-correlation kernel to zero and evidently is 
badly in error even at low $r_s$\cite{21}. 

Turning to the results shown in the top panel of Figure 1, a pleasant surprise in 2D is that 
ALDA does not overestimate badly the correlation energy as was shown by Lein {\it et al.}\cite{8} to be 
the case in 3D. In fact, in the range $2 < r_s < 10$ the ALDA results are seen to be quite close to 
the DPGT results, in which the full nonlocality of the static kernel in space has been included. 
The reason for this can be inferred from the work of Davoudi {\it et al.}\cite{22}, who showed that in 
2D satisfying the compressibility sum rule reproduces quite accurately the $q$-dependence of the 
static kernel almost up to $q \simeq 2k_F$. Following Lein {\it et al.}\cite{8} we have also examined 
frequency-dependent forms of the long-wavelength kernel and in particular that proposed by Qian 
and Vignale\cite{23}. We have found that this overestimates the correlation energy by as much as $\simeq 0.1$ Ryd 
at $r_s=3$. It appears, therefore, that nonlocality of the kernel in space should be included if 
its nonlocality in time is accounted for in the calculation of the correlation energy. 

This is, of course, what is being done in the AKA calculation. The average deviation of 
its results in the top panel of Figure 1 is approximately $0.02$ Ryd, which 
is somewhat better than in ALDA or in DPGT. Taking $g(0)$ 
in the AKA kernel from the ladder-diagram calculation\cite{19} 
raises this deviation to about $0.03$ Ryd and, most importantly, dropping the frequency 
dependence of the AKA kernel brings it to about $0.04$ Ryd. The summary conclusion of all 
these calculations is, therefore, that as far as the 
correlation energy in 2D is concerned one does 
already fairly well by taking a constant exchange-correlation kernel 
adjusted to the compressibility sum rule and that nonlocality of the kernel 
in both space and time needs including in order to obtain some improvement in the results. 

Figure 1 in its top panel further shows that STLS 
gives a very good estimate of the 
correlation energy, with a deviation of about $0.004$ Ryd on average. 
In this case inclusion of the 
frequency dependence of the kernel by means of the QSTLS recipe has only a minor effect, 
bringing the average deviation down to about $0.0035$. Consistency of the exchange-correlation 
kernel with the electron-pair structure thus appears to be important, as far as the calculation of 
the correlation energy is concerned. This fact may give a useful suggestion for further 
improvement of the more sophisticated kernels that have 
been proposed in the more recent work. We may also remark that Dobson {\it et al.}\cite{24} 
have recently extended the STLS scheme to 
inhomogeneous electronic systems and shown that it gives good results for the correlation 
energy of jellium slabs of finite thickness, lying within $3\%$ of the Diffusion Monte Carlo 
results of Acioli and Ceperley\cite{25}.
	
We turn now to Figures 2-4, in which following again Lein {\it et al.}\cite{8} we provide an 
analysis of the correlation energy at $r_s=1$ into contributions from correlations between density 
fluctuations of different wave vectors and different imaginary frequencies. Equations (\ref{7}) and 
(\ref{8}) naturally define a wave-vector analysis $\varepsilon_c(q)$ if we write
\begin{equation}\label{22}
\varepsilon_c=\int_0^{\infty}\varepsilon_c(q)d (q/k_F)
\end{equation}
and an imaginary-frequency analysis $\varepsilon_c(u)$ if we write
\begin{equation}\label{23}
\varepsilon_c=\int_0^{\infty}\varepsilon_c(u)d (2 m u/\hbar k^2_F)\,.
\end{equation}
Gori-Giorgi {\it et al.}\cite{26} have obtained an "exact" wave-vector analysis through the Fourier 
transform of the coupling-averaged correlation-hole density and built an analytic model for it by 
combining exactly known limiting behaviors with Diffusion Monte Carlo data, as in the work of 
Perdew and Wang\cite{27} on the 3D EG.

Figure 2 compares the results for $\varepsilon_c(q)$ given by the theories of present interest with the 
QMC ones that have kindly been communicated to us by Dr. P. Gori-Giorgi\cite{26}. While all 
theories perform similarly well in this test, it is pleasant to notice the reasonably good behavior 
shown by the ALDA. Finally, the imaginary-frequency analysis of the various theories is 
reported in Figures 3 and 4, in the low-$u$ and large-$u$ regimes respectively. As in the 3D case\cite{8}, 
in all theories $\varepsilon_c(u)$ starts with a finite negative value at $u=0$ and vanishes at large $u$. 
In ALDA $\varepsilon_c(u)$  becomes positive at $u \simeq 7 \hbar k_F^2/(2m)$ and ultimately 
vanishes from above.

To conclude our work, in Appendix A we briefly comment on how the coupling-constant 
dependence of the frequency-dependent exchange-correlation kernel in an inhomogeneous 
electronic system of arbitrary dimensionality may be found from the knowledge of its density 
dependence at full coupling strength.

\begin{acknowledgements}
This work was partially supported by MIUR through the PRIN2001 Initiative. One of us 
(M. P.) acknowledges partial support from Fondazione Angelo Della Riccia (Firenze) and from 
the National Science Foundation under Grant DMR 0115974. M. P. T. wishes to thank the 
Condensed Matter Group of the Abdus Salam International Center for Theoretical Physics for 
their hospitality during the preparation of the manuscript.  
We are grateful to Dr. G. S. Atwal for communicating to us Eqs. (19) and (20) and to Dr. P. Gori-Giorgi for 
sending to us prior to publication the "exact" results shown in Figure 2.
\end{acknowledgements}

\section*{APPENDIX A: SCALING OF THE 
INHOMOGENEOUS EXCHANGE-CORRELATION KERNEL}

From time-dependent density functional theory and dynamical 
scaling Lein {\it et al.}\cite{8} have derived a coordinate-scaling relation in dimensionality D=3
for the coupling-constant dependence of the exchange-correlation kernel $f_{\rm xc}[n]({\bf r}, {\bf r}';\omega)$ 
of an inhomogeneous many-electron system.
	
It is easy to generalize their proof to an inhomogeneous system in arbitrary space 
dimensionality D: namely, it is possible to prove that
\begin{equation}\label{24}
f^{\lambda}_{\rm xc}[n]({\bf r}, {\bf r}';\omega)=\lambda^2 f_{\rm xc}[n'](\lambda{\bf r},\lambda{\bf r}';\omega/\lambda^2)
\end{equation}
where $n'({\bf r},t)=\lambda^{-{\rm D}}n({\bf r}/\lambda,t/\lambda^2)$. 
In fact, the dimensionality of the system enters the proof 
only insofar as integrations of the time-dependent 
density $n({\bf r},t)$ over space are involved.

\newpage
\begin{figure}
\begin{center}
\includegraphics[scale=0.6]{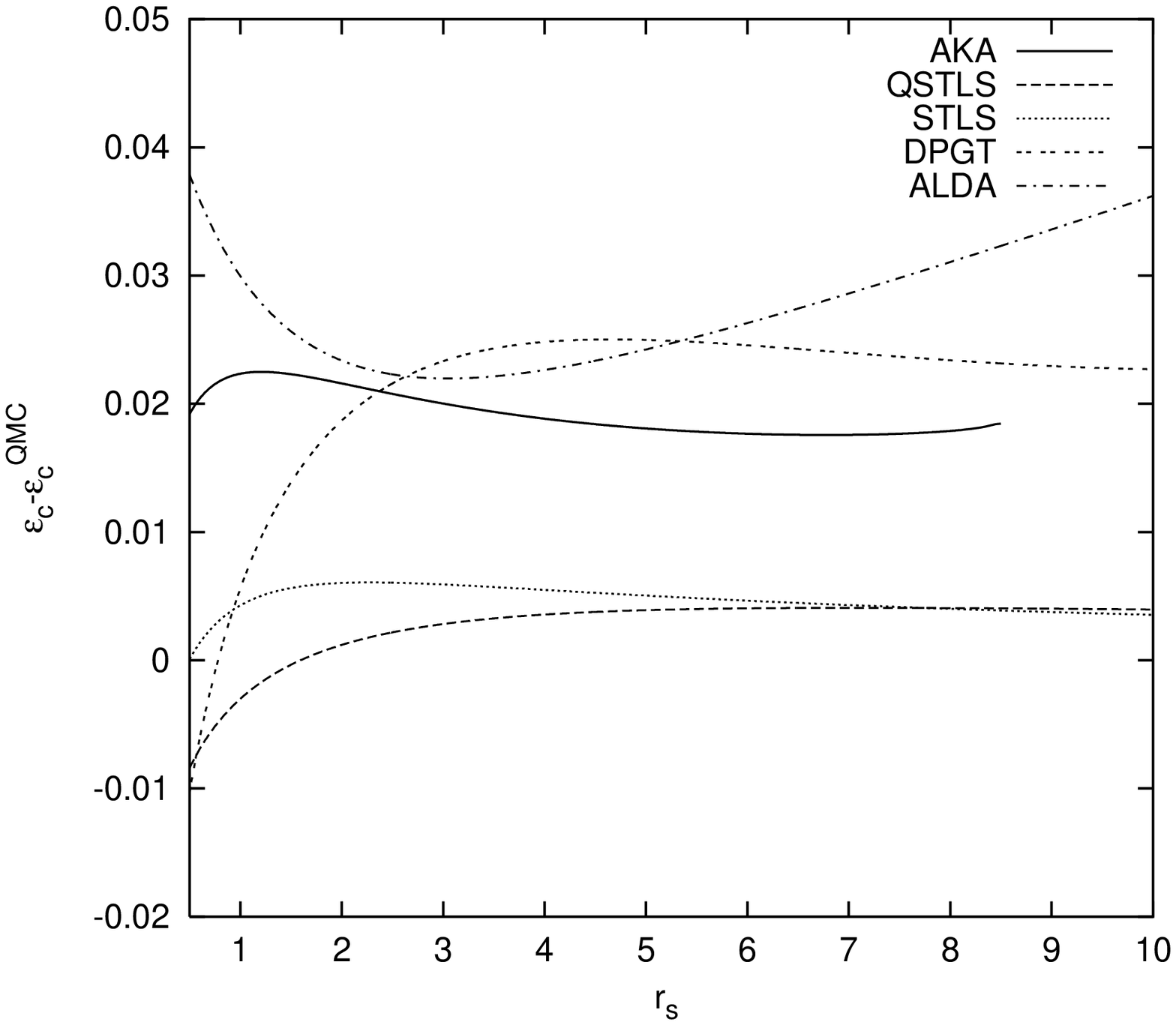}
\includegraphics[scale=0.6]{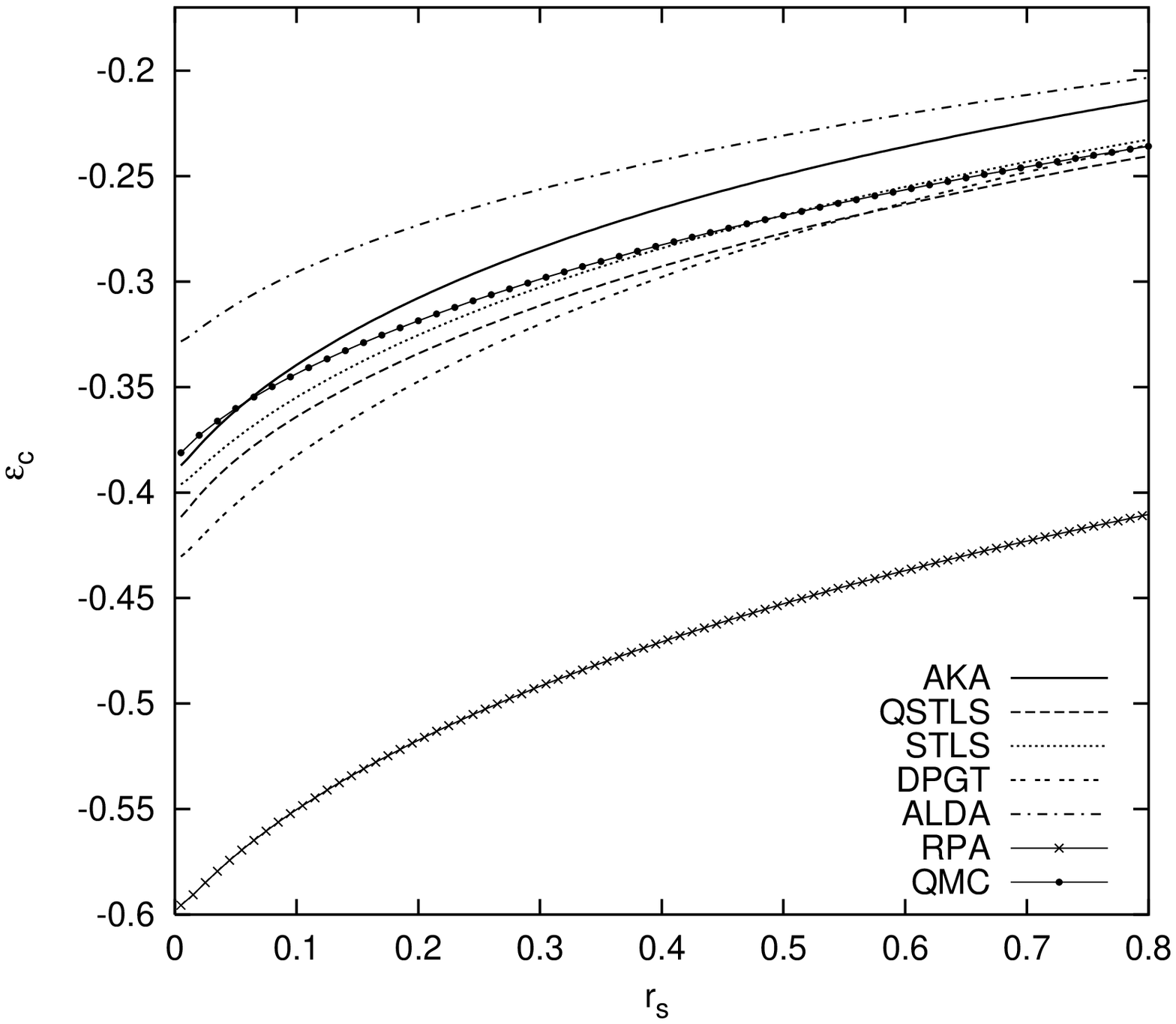}
\caption{Top panel: deviation of approximate correlation energy from the QMC correlation energy of Rapisarda and Senatore (in units of Ryd) as a function of $r_s$ for $0.5 \leq r_s \leq 10$ 
(the AKA results are omitted for $r_s 8$ where they do not seem to vary smoothly). 
Bottom panel: approximate and QMC correlation energy (in units of Ryd) as a function of $r_s$ for $0 \leq r_s \leq 0.8$ (the QMC results also embody the limiting form in Eq. (21)).}
\end{center}
\label{f1}
\end{figure}

\begin{figure}
\begin{center}
\includegraphics[scale=0.6]{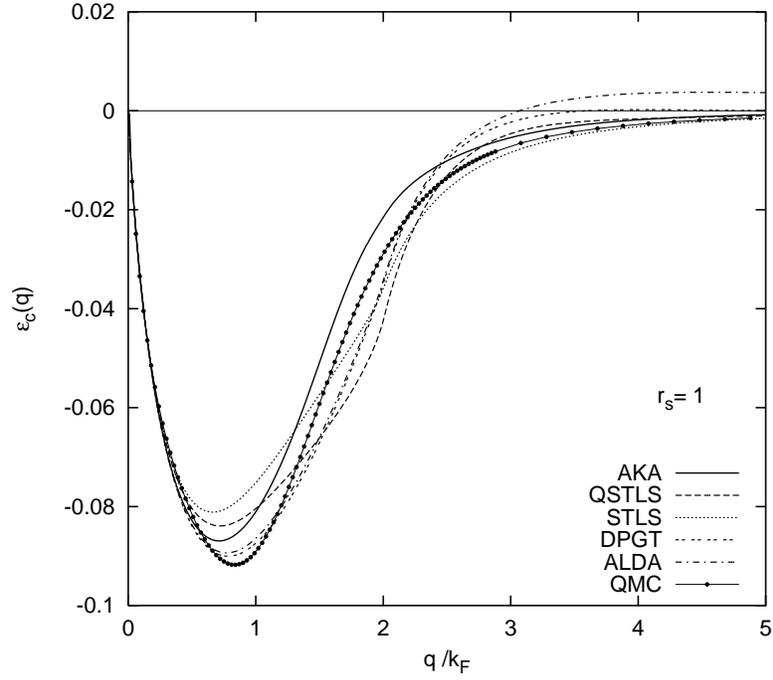}
\caption{Wave-vector analysis of the correlation energy per electron (in units of Ryd) at $r_s=1$ as a function of $q/k_F$.}
\end{center}
\label{f2}
\end{figure}

\begin{figure}
\begin{center}
\includegraphics[scale=0.6]{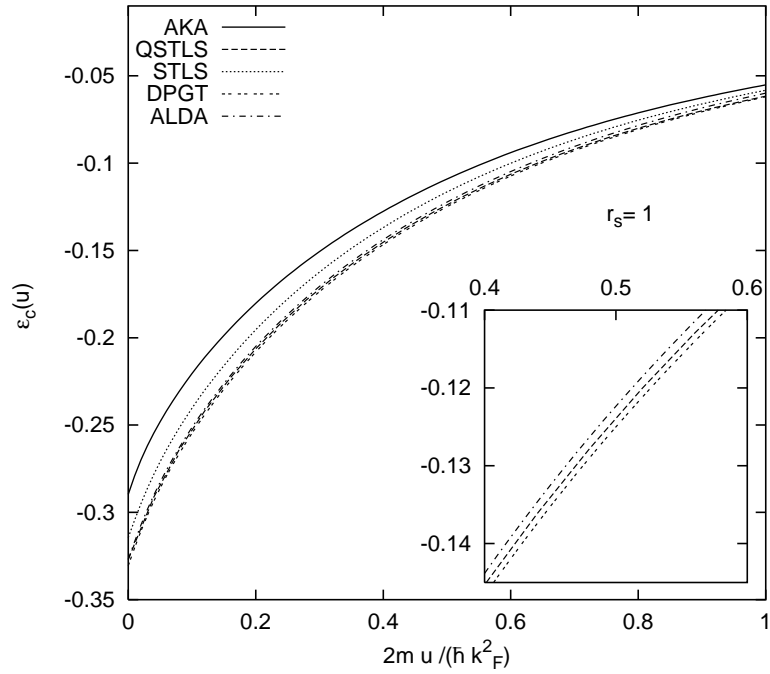}
\caption{Imaginary-frequency analysis of the correlation energy per electron (in units of Ryd) at $r_s=1$ as a function of $ 2m u/(\hbar k_F^2)$ in the low-$u$ regime. In the inset an enlargment of the results from ALDA, QSTLS and DPGT is shown.}
\end{center}
\label{f3}
\end{figure}

\begin{figure}
\begin{center}
\includegraphics[scale=0.6]{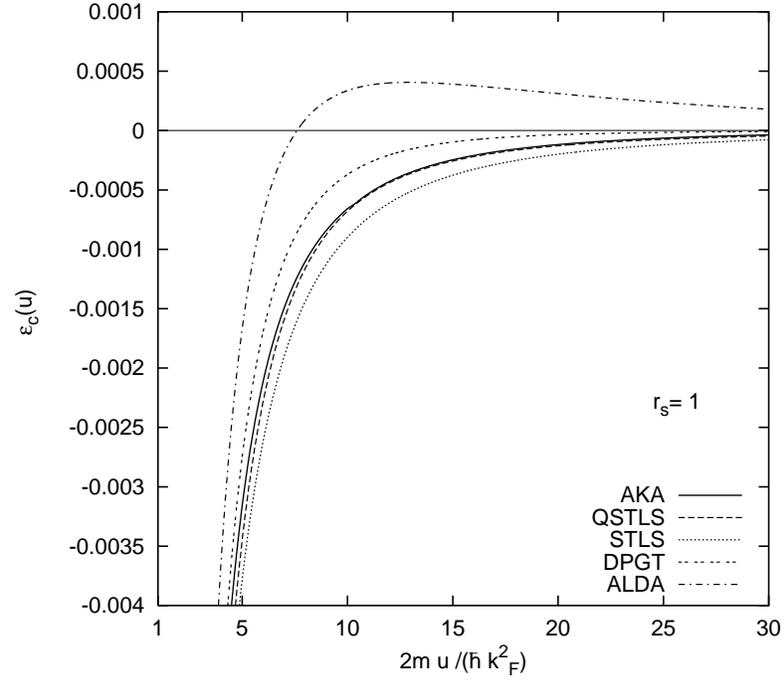}
\caption{Imaginary-frequency analysis of the correlation energy per electron (in units of Ryd) at $r_s=1$ as a function of $2m u/(\hbar k_F^2)$ in the large-$u$ regime.}
\end{center}
\label{f4}
\end{figure}

\end{document}